\documentclass[preprint,superscriptaddress,floatfix,prl,amssymb,amsmath]{revtex4}
\usepackage[dvips]{graphicx}
\usepackage{epsfig}
\usepackage{tabularx}
\usepackage{pifont}
\usepackage{color}
\usepackage[pdfpagemode=UseNone,colorlinks=true,linkcolor=blue,citecolor=blue]{hyperref}

\newcommand{\bh}{\hat{b}}
\newcommand{\bd}{\hat{b}^\dagger}
\newcommand{\nm}{\bar{n}}
\newcommand{\nth}{\bar{n}_{\mathrm{th}}}

\newcommand{\Geff}{\Gamma_{\mathrm{eff}}}
\newcommand{\Gopt}{\Gamma_{\mathrm{opt}}}
\newcommand{\Gm}{\Gamma_{\mathrm{m}}}

\newcommand{\Omegam}{\Omega_{\mathrm{m}}}

\newcommand{\Temp}{T_{\mathrm{bath}}}

\newcommand{\Wm}{\Omega_{\mathrm{m}}}

\newcommand{\Weff}{\Omega_{\mathrm{eff}}}
\newcommand{\Wexc}{\Omega_{\mathrm{exc}}}
\newcommand{\ud}{\mathrm{d}}

\newcommand{\dfm}{\delta f_{\mathrm{m}}}

\begin{document}

%\sloppy

\title{Probing quantum gravity effects with quantum mechanical oscillators}

\author{M. Bonaldi}
\affiliation{Institute of Materials for Electronics and Magnetism, Nanoscience-Trento-FBK Division, I-38123 Povo, Trento, Italy}  
\affiliation{Istituto Nazionale di Fisica Nucleare (INFN), Trento Institute for Fundamental Physics and Application, I-38123 Povo, Trento, Italy }
\author{A. Borrielli}
\affiliation{Institute of Materials for Electronics and Magnetism, Nanoscience-Trento-FBK Division, I-38123 Povo, Trento, Italy}  
\affiliation{Istituto Nazionale di Fisica Nucleare (INFN), Trento Institute for Fundamental Physics and Application, I-38123 Povo, Trento, Italy }
\author{A. Chowdhury}
\affiliation{CNR-INO, L.go E. Fermi 6, I-50125 Firenze, Italy }
\affiliation{INFN, Sezione di Firenze, Via Sansone 1, I-50019 Sesto Fiorentino (FI), Italy}
\author{G. Di Giuseppe}
\affiliation{School of Science and Technology, Physics Division, University of Camerino, via Madonna delle Carceri, 9, I-62032 Camerino (MC), Italy }
\affiliation{INFN, Sezione di Perugia, via A. Pascoli, I-06123 Perugia, Italy}
\author{W. Li}
\affiliation{School of Science and Technology, Physics Division, University of Camerino, via Madonna delle Carceri, 9, I-62032 Camerino (MC), Italy }
\author{N. Malossi}
\affiliation{School of Science and Technology, Physics Division, University of Camerino, via Madonna delle Carceri, 9, I-62032 Camerino (MC), Italy }
\author{F. Marino}
\affiliation{CNR-INO, L.go E. Fermi 6, I-50125 Firenze, Italy }
\affiliation{INFN, Sezione di Firenze, Via Sansone 1, I-50019 Sesto Fiorentino (FI), Italy}
\author{B. Morana}
\affiliation{Dept. of Microelectronics and Computer Engineering /ECTM/DIMES, Delft University of Technology, Feldmanweg 17, 2628 CT  Delft, The Netherlands }
\author{R. Natali}
\affiliation{School of Science and Technology, Physics Division, University of Camerino, via Madonna delle Carceri, 9, I-62032 Camerino (MC), Italy }
\affiliation{INFN, Sezione di Perugia, via A. Pascoli, I-06123 Perugia, Italy}
\author{P. Piergentili}
\affiliation{School of Science and Technology, Physics Division, University of Camerino, via Madonna delle Carceri, 9, I-62032 Camerino (MC), Italy }
\affiliation{INFN, Sezione di Perugia, via A. Pascoli, I-06123 Perugia, Italy}
\author{G. A. Prodi}
\affiliation{Istituto Nazionale di Fisica Nucleare (INFN), Trento Institute for Fundamental Physics and Application, I-38123 Povo, Trento, Italy }
\affiliation{Dipartimento di Fisica, Universit\`a di Trento, I-38123 Povo, Trento, Italy}
\author{P. M. Sarro}
\affiliation{Dept. of Microelectronics and Computer Engineering /ECTM/DIMES, Delft University of Technology, Feldmanweg 17, 2628 CT  Delft, The Netherlands }
\author{E. Serra}
\affiliation{Institute of Materials for Electronics and Magnetism, Nanoscience-Trento-FBK Division, I-38123 Povo, Trento, Italy}  
\affiliation{Istituto Nazionale di Fisica Nucleare (INFN), Trento Institute for Fundamental Physics and Application, I-38123 Povo, Trento, Italy }
\affiliation{Dept. of Microelectronics and Computer Engineering /ECTM/DIMES, Delft University of Technology, Feldmanweg 17, 2628 CT  Delft, The Netherlands }
\author{P. Vezio}
\affiliation{INFN, Sezione di Firenze, Via Sansone 1, I-50019 Sesto Fiorentino (FI), Italy}
\affiliation{European Laboratory for Non-Linear Spectroscopy (LENS), Via N. Carrara 1, I-50019 Sesto Fiorentino (FI), Italy }
\author{D. Vitali}
\affiliation{CNR-INO, L.go E. Fermi 6, I-50125 Firenze, Italy }
\affiliation{School of Science and Technology, Physics Division, University of Camerino, via Madonna delle Carceri, 9, I-62032 Camerino (MC), Italy }
\affiliation{INFN, Sezione di Perugia, via A. Pascoli, I-06123 Perugia, Italy}
\author{F. Marin}
\affiliation{CNR-INO, L.go E. Fermi 6, I-50125 Firenze, Italy }
\affiliation{INFN, Sezione di Firenze, Via Sansone 1, I-50019 Sesto Fiorentino (FI), Italy}
\affiliation{European Laboratory for Non-Linear Spectroscopy (LENS), Via N. Carrara 1, I-50019 Sesto Fiorentino (FI), Italy }
\affiliation{Dipartimento di Fisica e Astronomia, Universit\`a di Firenze, Via G. Sansone 1, I-50019 Sesto Fiorentino (FI), Italy  }

\date{\today}
\begin{abstract}
Phenomenological models aiming to join gravity and quantum mechanics often predict effects that are potentially measurable in refined low-energy experiments. For instance, modified commutation relations between position and momentum, that accounts for a minimal scale length, yield a dynamics that can be codified in additional Hamiltonian terms. When applied to the paradigmatic case of a mechanical oscillator, such terms, at the lowest order in the deformation parameter, introduce a weak intrinsic nonlinearity and, consequently, deviations from the classical trajectory. This point of view has stimulated several experimental proposals and realizations, leading to meaningful upper limits to the deformation parameter. All such experiments are based on classical mechanical oscillators, i.e., excited from a thermal state. We remark indeed that  decoherence, that plays a major role in distinguishing the classical from the quantum behavior of (macroscopic) systems,  is not usually included in phenomenological quantum gravity models. However, it would not be surprising if peculiar features that are predicted by considering the joined roles of gravity and quantum physics should manifest themselves just on purely quantum objects. On the base of this consideration, we propose experiments aiming to observe possible quantum gravity effects on macroscopic mechanical oscillators that are preliminary prepared in a high purity state, and we report on the status of their realization.
\end{abstract}

\maketitle

\section{Introduction}
\label{intro}

The search for a theory of Quantum Gravity (QG) has received so far little guidance from experimental observations. 
Quantum gravitational effects are indeed expected to emerge at the Planck-scale, i.e., at extremely small distances $L_p =\sqrt{\hbar G/c^3} = 1.6 \times 10^{-35}$ m and/or high energies $E_p=1.2 \times 10^{19}$GeV, well beyond the possibilities of any current and foreseeable accelerator. In this context, high-energy astronomical events have been considered as the privileged arena to unveil Planck-scale effects \cite{amelino-camelia2,jacob,tamburini}. For instance, accurate measurements of the arrival time of $\gamma$-ray photons propagated over cosmological distances allowed to place stringent limits to the predicted variations of light speed with energy \cite{abdo}. Further QG phenomena could be tested also thanks to current and future gravitational-wave (GW) observations \cite{gw}. Nevertheless, the approach based on astronomical observations inherently suffers from the lack of control of the experimental conditions and, in some cases, from a limited knowledge of the underlying physical mechanisms.
 
The situation has greatly improved in the last years thanks to a number of investigations that, starting from existing QG candidate theories, derived a number of possible low-energy signatures of Planck-scale physics. Relevant studies include, e.g., tests of quantum decoherence and state collapse models \cite{bassi}, QG imprints on initial cosmological perturbations \cite{weinberg}, cosmological variation of coupling constants \cite{damour}, TeV black holes in large extra-dimensions \cite{bleicher}, Planck-scale spacetime fuzziness \cite{amelinonat99} and generalized uncertainty principles (GUPs) \cite{veneziano,gross,garay,maggiore1,scardigli,jizba,ali2,Hoss2012}. In this article we focus on this latter class of models.

A minimal observable length is a common feature of QG theories and is usually assumed to be related to the quantum fluctuations of background spacetime metric. Such fluctuations are expected to introduce an additional quantum uncertainty in position measurement. The Heisenberg uncertainty principle does not set limits on the accuracy in the measurement of a given observable: an arbitrarily precise measurement of the position of a particle is indeed possible at the cost of our knowledge about its momentum. A generalization of the uncertainty relation is thus required to incorporate the existence of such a minimal measurable length. In its most common form, a GUP between position and momentum is written as 
\begin{equation}
\Delta x \Delta p \geq \frac{\hbar}{2} \left[1+\beta_0 \left(\frac{L_p \Delta p}{\hbar}\right)^2\right]  \quad .
\label{eq1}
\end{equation}
Eq. (\ref{eq1}) implies a nonzero minimal uncertainty in position measurements $\Delta x_{min} = \sqrt{\beta_0} L_p$. The dimensionless parameter $\beta_0$ is usually assumed to be around unity, in which case the corrections are negligible unless energies (lengths) are close to the Planck energy (length). Any experimental bound $\beta_0 > 1$ thus set a new physical length scale $\sqrt{\beta_0} L_p$ below which Planckian corrections to quantum mechanics could become significant \cite{das}. Meaningful limits should fall below the electroweak scale located at $10^{17} L_p$, where deviations from standard theory have been already ruled out. 
As a further consequence of Eq. (\ref{eq1}), the ground state energy $E_{min}$ of a quantum harmonic oscillator with frequency $\omega_0$ is larger with respect to the standard $\hbar \omega_0/2$. Therefore, an experiment measuring a minimal modal energy $E_{exp}$ provides an upper limit to the corresponding $E_{min}$. In Refs. \cite{auriga1,auriga2} the sub-mK cooling of the first longitudinal mode of the AURIGA bar detector is considered. Although the system was still far from its quantum ground state, an upper bound below the electroweak scale was obtained thanks to the huge modal mass of the bar.

The GUP described by Eq. (\ref{eq1}) can be associated to the deformed commutation relation \cite{kempf}
\begin{equation}
[x,p] = i \hbar \left[1+\beta_0 \left(\frac{L_{\mathrm{P}} \, p}{\hbar}\right)^2\right] \, .  
\label{eq2}
\end{equation}
Eq. (\ref{eq2}) implies changes in the whole energy spectrum of quantum systems, as well as in the time evolution of a given observable. Corrections to Landau levels and Lamb shift associated to modified quantum Hamiltonians have been computed in Ref. \cite{das}. The modified spectrum of a quantum harmonic oscillator has been calculated in Refs. \cite{kempf,chang,lewis}. Expressions for generalized coherent states has been obtained in Refs. \cite{ching,pedram1} and the modified time evolution and expectation values of position and momentum operators are discussed in Refs. \cite{nozari1,nozari2,pedram2}. 
First estimates based on accurate measurements of the Lamb shift \cite{das} and the 1S-2S level energy difference in hydrogen atom \cite{quesne} led to upper bounds slightly larger than the electroweak scale. Much stronger constraints from simple atomic systems are not straightforward, due to the small masses involved.

In this respect, cavity opto-mechanical systems \cite{AspelRMP} in which micro- and nano-mechanical resonators can be coherently coupled with optical and microwave cavity fields, offer an alternative route to astrophysical and spectroscopic measurements. The optomechanical coupling allows both to prepare macroscopic mechanical oscillators into massive quantum states and to monitor their motion by means of highly-sensitive interferometric techniques.
Recent experiments achieved the cooling of macroscopic oscillators down to thermal occupation numbers below unity \cite{meenehan,riedinger,Peterson2016}, as well as the preparation of mechanical squeezed states \cite{Wollman2015,Pirkkalainen2015,Lecocq2015,Lei2016,Chowdhury2020}. Non-classical signatures of the motion have been observed around one or few normal modes of the mechanical oscillator \cite{Peterson2016,Chowdhury2020,Safavi2012,Purdy2015,Underwood2015,Sudhir2017a}. Each mode is associated to a massive degree of freedom and actually behaves as a quantum harmonic oscillator.

An optomechanical scheme for a direct measurement of the canonical commutator of the center of mass of a massive object was proposed in Ref. \cite{pikovski} and further developed in Refs. \cite{bosso} and \cite{plenio}. 
Further studies focused on the dynamical consequences of the modified commutator of Eq. (\ref{eq2}). In this framework, Bawaj {\it et al.} \cite{bawaj} underline that the corresponding Heisenberg equations become nonlinear and the time evolution of the position operator exhibits a third harmonic term and a dependence of the oscillation frequency on its amplitude. The strength of such effects depends on the deformation parameter $\beta_0$. Experiments proposed in Ref. \cite{bawaj} were realized by the same authors exploiting macroscopic nano- and micro-oscillators with masses ranging from $10^{-11}$ to 
$10^{-5}\,$kg, and more recently by Bushev {\it et al.} \cite{tobar} with a sub-kilogram sapphire oscillator. These experiments have lowered previous limits on $\beta_0$ by several orders of magnitude. 

Nevertheless, all the above experiments were operated in a classical regime
in which the deformation parameter has a measurable effect on the dynamics
of the expectation values of the mechanical resonator, with the quantum
fluctuations playing no relevant role since they are overwhelmed by
classical noise of thermal origin.

It is instead interesting and unexplored up to now to repeat the same class
of experiments in a quantum regime in which the resonator remains in a high
purity state, and therefore describable in terms of a wavefunction. In this
regime, the effect of quantum fluctuations on the expectation value
dynamics cannot be neglected anymore and the result of the experiment would
provide novel useful hints for the development of a theory able to consider
gravitational phenomena within a consistent quantum treatment. Due to the
novel quantum regime, even if the bounds on the deformation parameter that can be achieved are probably not
as good as those of Refs. \cite{bawaj,tobar}, the results acquire a particular interest.

\section{Experimental schemes and protocols}
\label{sec2}

Opto-mechanical techniques allow to prepare a macroscopic mechanical oscillator close to its ground state. The transition between a ``classical'' and a ``quantum'' state is not sharp, and its simplest and most immediate quantification uses the purity of the state. This indicator is defined as the trace of the square of the density matrix, which is equal to one for a pure state. For a thermal oscillator, the purity can be written as $1/(1+2 \nm)$, where $\nm$ is the mean phonon occupation number. A translation in the phase space creates a coherent state with a unitary transformation, that preserves the state purity while increasing the amplitude of the motion. 

Optical cooling achieves the goal of a low $\nm$, in spite of the relatively high background temperature, by coupling the mechanical oscillator to the photon bath that compete with the phononic background reservoir. The opto-mechanical coupling Hamiltonian is non-dissipative, therefore one can argue that a meaningful investigation of possible phenomena related to modifications of the standard quantum mechanical scenario can be performed even maintaining a stationary optical cooling. On the other hand, it is also true that an efficient cooling requires a strong coupling of the mechanical oscillator to the intracavity optical field, that is actually coupled to the outside world and potentially read out. A more conservative scheme would therefore imply the analysis of the oscillator properties with the weakest possible probe, in the absence of strong coupling, in a short period after the preparation of the high purity state and before re-thermalization.

\begin{figure}[htb!]
\begin{center}
   {\includegraphics[width=.45\textwidth]{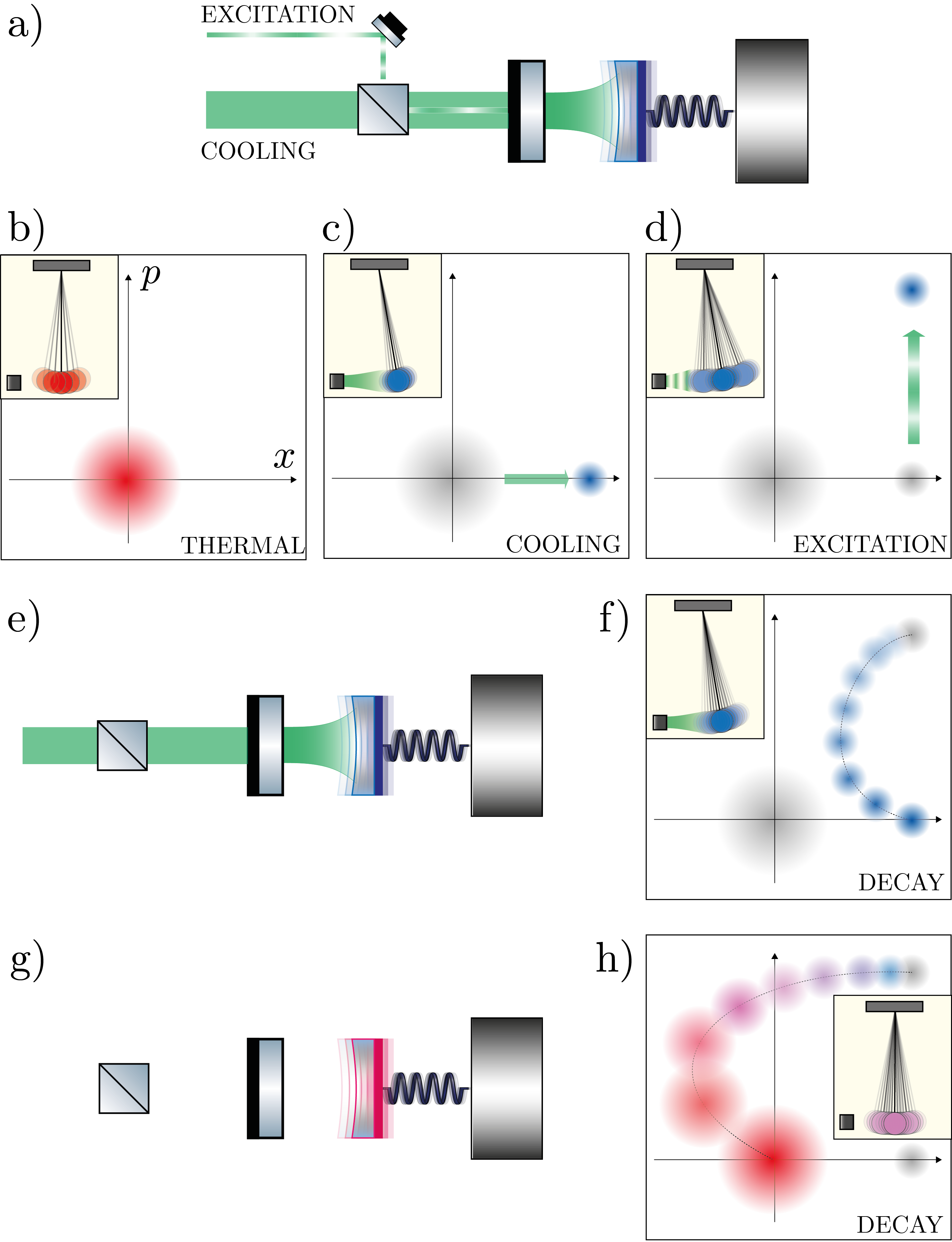}}
 \caption{{\bf a)} State preparation: optical cooling and coherent excitation of the mechanical oscillator are achieved by means of laser beams, one of which is amplitude modulated.
 {\bf b--d)} Phase space representations of the state of the mechanical oscillator in an initial thermal state (b), optically cooled to a low phonon-number state (c), and finally displaced in the phase space by a coherent modulation. {\bf e)} Sketch of the first experimental proposal: the decay dynamics of the mechanical oscillator is detected once \emph{only} the excitation beam is turned off. {\bf f)} Phase space representation of the decay dynamics for the first proposal. {\bf g)} Sketch of the second experimental proposal: the decay dynamics of the mechanical oscillator is detected once \emph{both} the cooling and excitation beams are turned off. {\bf h)} Phase space representation of the decay dynamics for the second proposal.}
\label{fig_protocolli}
\end{center}
\end{figure}

Based on these considerations, we propose two different experimental protocols, synthesized in Fig. (\ref{fig_protocolli}). The first scheme includes a preliminary state preparation by optical cooling, followed by an experimental cycle where the oscillator is firstly excited (i.e., prepared in a coherent state) [Fig. (\ref{fig_protocolli}a)], then observed by a probe during the relaxation that follows the switching off of the excitation signal [Fig. (\ref{fig_protocolli}e)]. The analysis is here quite close to the one described in Ref. \cite{bawaj}: the most evident effect of the dynamics with a deformed commutator is a frequency shift having a quadratic dependence on the oscillation amplitude. This shift is measured during the oscillator decay, exploiting the exponential decrease of the amplitude. In the present scheme we have however to pay attention to at least three additional aspects: a) due to the optical damping, the decay time is much shorter than the natural oscillator relaxation time, b) the optical spring realized by the intracavity laser radiation (cooling and probe fields) must be carefully controlled to avoid spurious and deleterious effects on the frequency stability, and c) the optical cavity, with a strong intracavity field, introduces additional nonlinearities that must be accurately considered and evaluated before extracting from the data a meaningful estimate of the deformation parameter or, more likely, an upper limit to such constant.   

The second scheme implements a cycle in three steps: the optical cooling and the excitation [Fig. (\ref{fig_protocolli}a)] are followed by the readout with a weak probe after the complete shutdown of the laser system used for cooling and excitation, in a short period [Fig. (\ref{fig_protocolli}g)]. A possible dependence of the oscillation frequency on its amplitude can be explored by varying the excitation strength, instead of relying on the decaying amplitude that would occur in a timescale longer than the re-thermalization time. To achieve a weak probe coupling, it can be useful to operate the probe laser field at a wavelength where the cavity Finesse is not too high. On the other hand, the cooling field is generally coupled to a high Finesse cavity for an optimized cooling efficiency. In this sense, one should find the optimal trade-off between strong coupling, giving a better signal-to-noise and lower statistical uncertainty, and weak coupling, the latter aiming to limiting both the disturbance on the interesting properties of the oscillator (i.e., approaching the ideal case of an isolated system), and the spurious optical spring effect and the consequent systematic errors.       

\section{Status of the experiment}
\label{sec3}

Our collaboration is working towards the implementation of the experimental schemes described in the previous Section, in two stages. The first is at cryogenic temperature (5-10 K), on a mechanical oscillator with a target phonon occupation number below 5 (thus, a purity above 0.1). For the second stage, we are preparing an ultra-cryogenic setup, achieving a base oscillator temperature below 1 K, with a target stationary mean phonon number below unity. 

The mechanical oscillator is a drum mode of a tensioned membrane. For this purpose, we have developed circular SiNx membranes, supported by a silicon ring frame. Diameter and thickness of the membranes are typically of the order of 1.5 mm and 100 nm respectively. The drum modes have eigenfrequencies starting from $\sim 200$ kHz and masses of the order of $10^{-10}$ kg. The silicon frame is suspended 
with alternating flexural and torsional springs, forming an on-chip ``loss shield'' structure \cite{Borrielli2014} that allows to achieve a mechanical quality factor $Q$ around $10^7$ for most of the drum modes. More information
about the design, fabrication and the characteristics of the device can be
found in the works by Borrielli {\it et al.} \cite{Borrielli2016} and Serra {\it et al.} \cite{Serra2016,Serra2018}. 

The membrane is embedded in a high Finesse optical cavity, forming an opto-mechanical setup that allows for optical cooling and readout of the oscillator motion. After the switching off of the cooling beam, the oscillator re-thermalizes  to the background temperature $\Temp$, and its mean phonon occupation number $\nm$ evolves according to 
\begin{eqnarray}
\nm(t) &=& \nm(0) \exp(-\Gm t) + \nth \left[1- \exp(-\Gm t)\right] \nonumber\\
&  \simeq & \nm(0) + \frac{k_{B}\Temp}{\hbar Q} t 
\label{nmt}
\end{eqnarray}
where $\Gm$ is the oscillator damping rate and $Q=\Omegam/\Gm$ its quality factor. The equilibrium thermal occupation number is $\nth = \left[\exp(\hbar \Omegam/k_{B}\Temp)-1\right]^{-1} \simeq k_{B}\Temp / \hbar \Omegam$ and the second equivalence in Eq. (\ref{nmt}) holds for short periods of time. The expected re-thermalization rate is 
of the order of one additional thermal phonon every $5 \mu$s for the cryogenic experiment, and one phonon every $100 \mu$s for the ultra-cryogenic stage.

In the following part of this Section we present a characterization of the setup working at cryogenic temperature, and a preliminary implementation of the experimental protocols that will allow the search of a dependence of the oscillation frequency from the amplitude, with a macroscopic mechanical resonator in a high purity state.

\begin{figure}
\resizebox{0.5\textwidth}{!}{\includegraphics{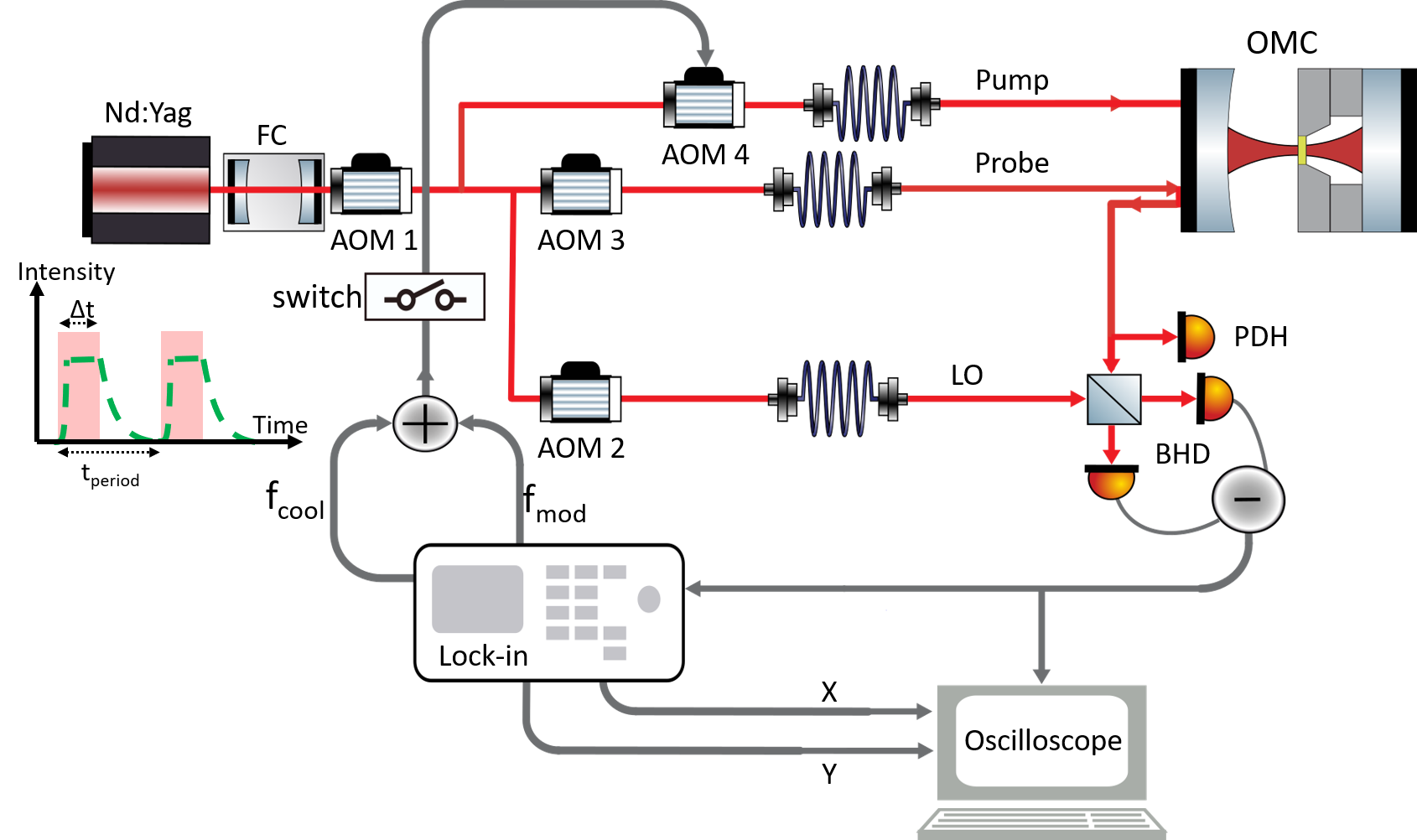}}
\resizebox{0.45\textwidth}{!}{\includegraphics{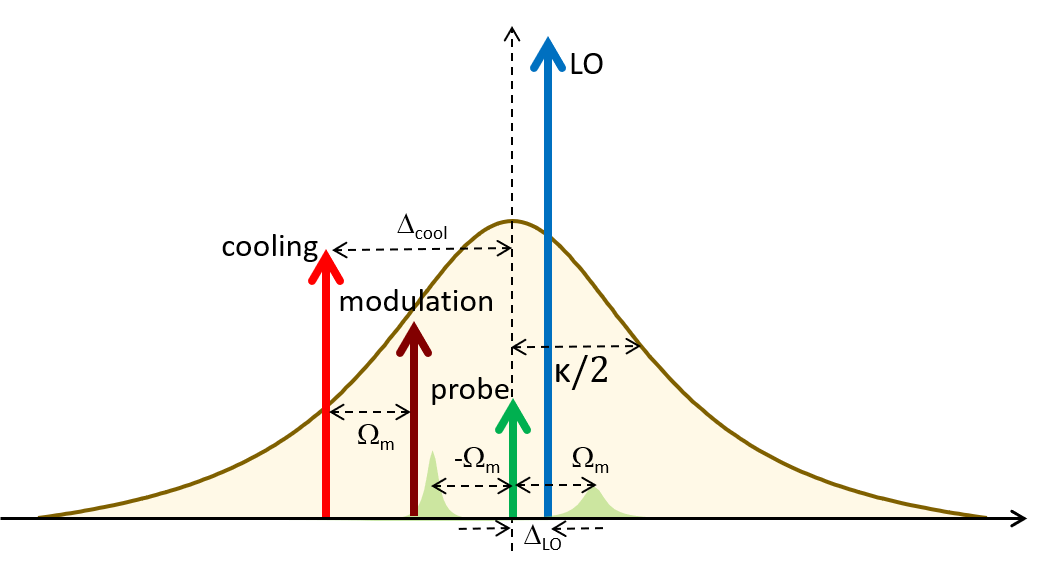}}
\caption{Layout of the experimental setup (see text) and conceptual scheme of the field frequencies. The LO is placed on the blue side of the probe and detuned by $\Delta_{\mathrm{LO}}  \ll  \Omegam$, therefore the Stokes line is on the red side of the LO, while the anti-Stokes line is on its blue side. In the heterodyne spectra, they are located respectively at $\Omegam+\Delta_{\mathrm{LO}}$ (Stokes) and $\Omegam-\Delta_{\mathrm{LO}}$ (anti-Stokes).}
\label{fig_setup}
\end{figure}

The scheme of the setup is shown in Fig. \ref{fig_setup}. The main laser source is a NPRO Nd:YAG laser, whose radiation is filtered by the transmission through a Fabry-Perot cavity (FC, having a linewidth of $66\,$kHz) and split into three beams, whose frequencies are controlled by means of acousto-optic modulators (AOM) and sent to the experimental bench by optical fibers. The probe beam is kept resonant with the optomechanical cavity (OMC) using a Pound-Drever-Hall detection (PDH) followed by a servo loop. Most of the reflected probe light ($\sim 10 \,\mu$W) is combined with the local oscillator (LO) beam ($\sim 2\,$mW) in a balanced detection (BHD). 
The LO frequency $\omega_{\mathrm{LO}}$ is shifted with respect to the probe by $\Delta_{\mathrm{LO}} = 2\pi \times 12\,$kHz, to realize a low-frequency heterodyne detection \cite{Pontin2018a}. The output of the BHD is both directly acquired and sent to a lock-in amplifier which demodulates the signal at $(\Wexc - 2\pi\times 4 \mathrm{kHz}) $, where $\Wexc$ is the excitation frequency. The two quadrature outputs of the lock-in are simultaneously acquired and off-line processed. 
The third beam (pump beam), orthogonally polarized with respect to the probe, is also sent to the cavity. Its field contains a main, cooling tone at a frequency $\omega_{\mathrm{cool}}$ 
red detuned from the cavity resonance by $\Delta_{\mathrm{cool}} = -2 \pi \times 700 \mathrm{kHz} $, 
and a modulation tone at a frequency $\omega_{\mathrm{cool}}+\Wexc$. The two tones are obtained by driving the AOM2 with the sum of two radiofrequency signals. 
All the radiofrequency sinusoidal signals used in the experiment, for driving the AOMs and as reference in the lock-in amplifier, are phase locked. 

The membrane oscillator is placed in a Fabry-Perot cavity of length $4.38$~mm, at $2$~mm from the cavity flat end mirror, forming a ``membrane-in-the-middle'' setup \cite{Jayich2008}. The cavity linewidth is $\kappa/2\pi = 2.1  \,\textrm{MHz}$. The optomechanical cavity is cooled in a helium flux cryostat, operating at 9 K during the measurements presented in this article. More details on the optomechanical system and its characterization are reported and discussed in Ref. \cite{Chowdhury2019}. In this work we exploit the (0,2) drum mode of the membrane at $\sim530\,$kHz, having a quality factor of $6.4 \times 10^6$ at cryogenic temperature (mechanical linewidth 0.08 Hz).  

The mechanical interaction of the oscillator with the resonant probe field produces in the latter motional sidebands around its main field frequency, displaced by $\pm \Omegam$, and proportional respectively to $\bh = X + i Y$ (anti-Stokes sideband) and $\bd = X-iY$ (Stokes sideband), where $X$ and $Y$ are the oscillator quadratures. In a thermal state, the variances of $\bh$ and $\bd$ are respectively $\langle \bd \, \bh \rangle  = \nth$ and $\langle \bh \, \bd \rangle  = \nth +1$ \cite{Clerk2010}. The interaction of the oscillator with the detuned intracavity radiation yields the following effects. The mechanical susceptibility is modified due to the position-dependent radiation pressure, giving a shifted resonance frequency $\Weff$ (optical spring effect) and a strong optical damping, broadening the resonance from the initial natural linewidth $\Gm$ to $\Geff$  \cite{AspelRMP,Arcizet2006}. As a consequence of such cold damping, the phonon occupation number is reduced from the initial value of $\nth$ to a much lower $\nm$ that includes the residual contribution of the thermal bath $(\nth \Gm/\Geff)$ and the effect of the back-action from the intracavity field. With an optically cooled oscillator, the ratio between Stokes and anti-Stokes sidebands can be written as $R = (\nm+1)/\nm$ and, for low enough $\nm$,
a deviation from unity of $R$ becomes measurable, providing a clear signature of the smooth transition between the classical motion and the quantum behavior and allowing a direct measurement of $\nm$. 

\begin{figure}
\resizebox{0.49\textwidth}{!}{\includegraphics{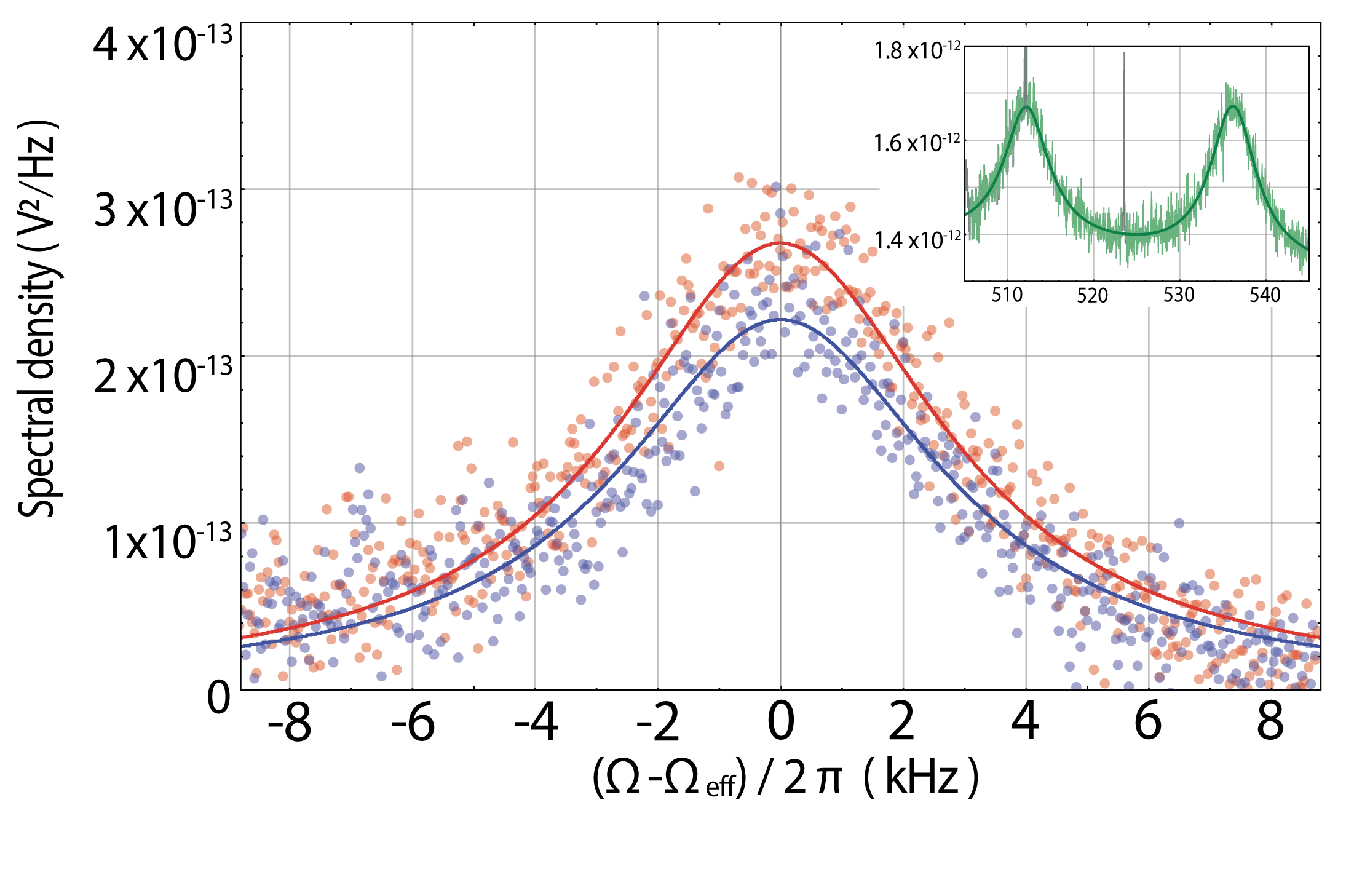}}
\caption{Spectra of the motional sidebands, measured at output of the BHD. The inset shows the original spectrum. The experimental signals corresponding to the two sidebands (dots), originally separated by 24 kHz, have been corrected for the residual probe detuning (see Ref. \cite{Chowdhury2019}) and superposed for an easier visual comparison, after subtraction of the background. Red: Stokes sideband. Blue: anti-Stokes sideband. Solid lines: Lorentzian fittings.}
\label{fig_sidebands}
\end{figure}

The two motional sidebands are distinguishable in the heterodyne spectrum where they appear separated by $24\,$kHz (see sketch in Fig. \ref{fig_setup}). An example of such sidebands, superimposed for a clearer visual comparison, is shown in Fig. (\ref{fig_sidebands}). The direct measurement of the sidebands ratio must be corrected for the residual probe detuning, as described in Ref. \cite{Chowdhury2019}. The analysis of the spectrum shown in the figure gives a phonon occupation number of $\nm = 5$, and an effective width of $\Geff = 2\pi \times 6\,$kHz. 

\begin{figure}
\resizebox{0.47\textwidth}{!}{\includegraphics{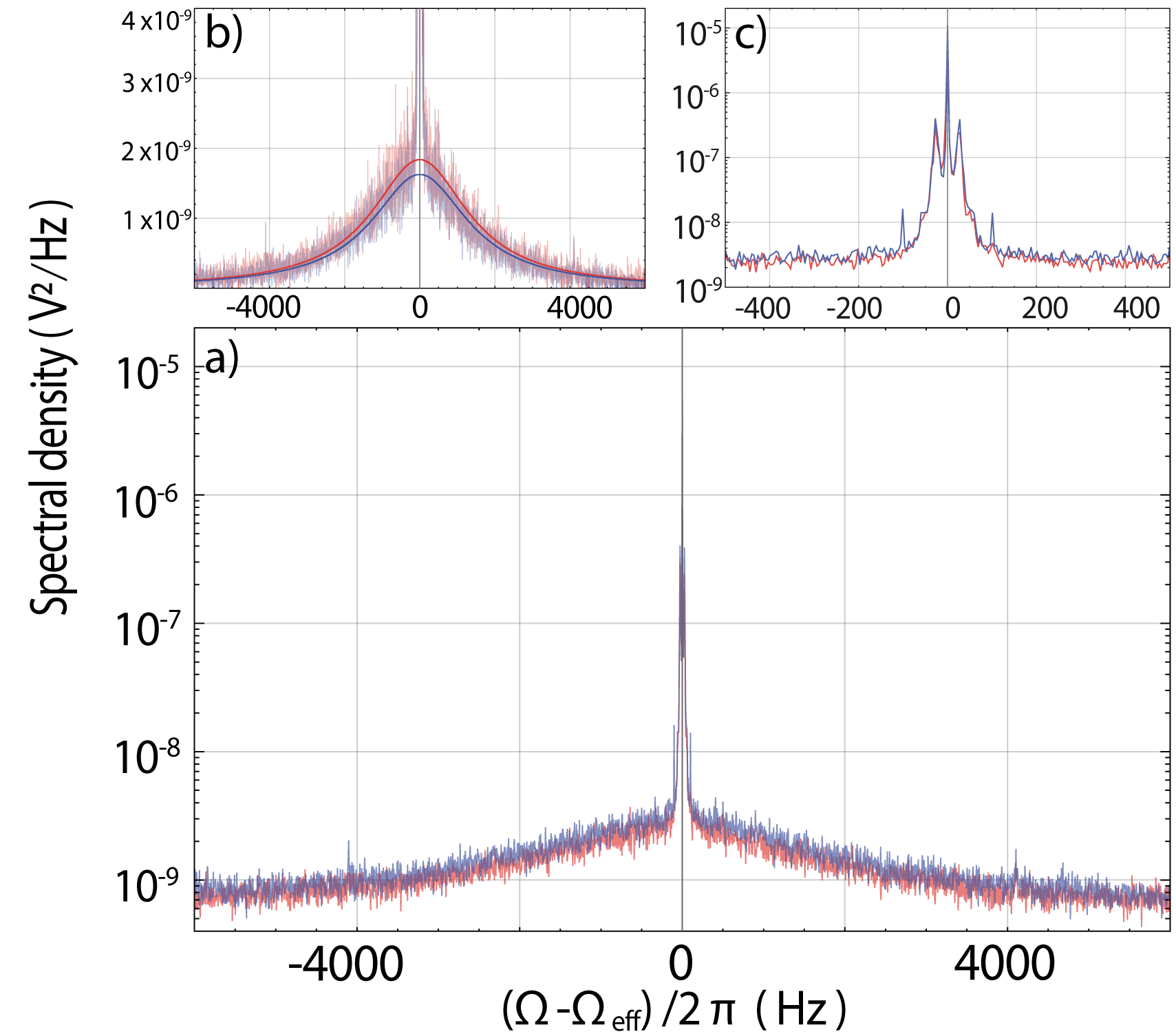}}
\caption{Spectra of the motional sidebands for a coherently excited oscillator. The excitation tone power is -60 dB lower than the cooling tone power, as measured at the input of AOM2. Red: Stokes sideband. Blue: anti-Stokes sideband. {\bf a)} Full spectra, superposed for a better visual comparison. {\bf b)} Spectra corrected as in Fig. (3) and fitted with Lorentzian shapes to deduce $\nm$. {\bf c)} Enlarged view around the coherent peak.}
\label{fig_excited}
\end{figure}

When the excitation tone is turned on, the modulated radiation pressure inside the cavity excites the oscillator with a coherent signal that adds to its quantum and thermal fluctuations. An example of the spectrum is shown in Fig. (\ref{fig_excited}). The amplitude of the coherent component $\alpha$ can be evaluated from the spectrum by considering that the ratio between the area of the narrow coherent peak and that of the Lorentzian background is $|\alpha|^2/(\nm+1/2)$. For the spectra in Fig. (\ref{fig_excited}) we calculate a phonon occupation number of $\nm = 6.6$ and a coherent amplitude $|\alpha|^2 = 35$. It is important to verify that the addition of the excitation tone do not significantly modify the oscillator noise spectrum. The experimental protocol is therefore a good approximation of an unitary operation, that maintains the oscillator state purity. An enlarged view around the coherent peak [Fig. (\ref{fig_excited}c)] shows indeed the presence of a weak pedestal, including minor peaks separated by multiples of the 50 Hz power line frequency. We attribute this effect to an up-conversion of the laser low-frequency noise caused by the modulation. However, in the shown spectra such additional fluctuations are limited in both amplitude and bandwidth (their timescale is much slower than the oscillator dynamic timescale $1/\Geff$), and they can be accounted, e.g., by considering a slowly varying coherent amplitude $\alpha(t)$. The situation is different if the intensity of the excitation tone is increased, until, for a ratio with the cooling tone of -30dB, we observe that the noise is completely dominated by the pedestal of the coherent peak. The experimentally achievable coherent amplitude range that maintains a negligible degradation of the oscillator state purity is still a matter of investigation. In the present work, we limit ourself to conservative values of excitation power.   

For the analysis of the oscillator weakly interacting with the probe field, we have implemented a periodic experimental cycle lasting 40 ms. For 30 ms the pump beam is on and the oscillator is simultaneously cooled and excited. This period is much longer than $1/\Geff$, therefore we can assume that the oscillator has reached a stationary condition before the end of this stage. The pump beam is subsequently turned off for a measurement period of 10 ms. This latter period, while much shorter than the free oscillator damping time $1/\Gm$, is still much longer than the re-thermalization time $\hbar Q/k_{B}\Temp$, and therefore it is sufficient to our purposes. The driver of the radiofrequency switch is kept synchronous with the AOM oscillators, such that the phase of the excitation sinusoidal force at the end of the 30 ms period just varies very slowly due to long-term drifts. We can therefore average the measured signal over many cycles. 

\begin{figure}
\resizebox{0.45\textwidth}{!}{\includegraphics{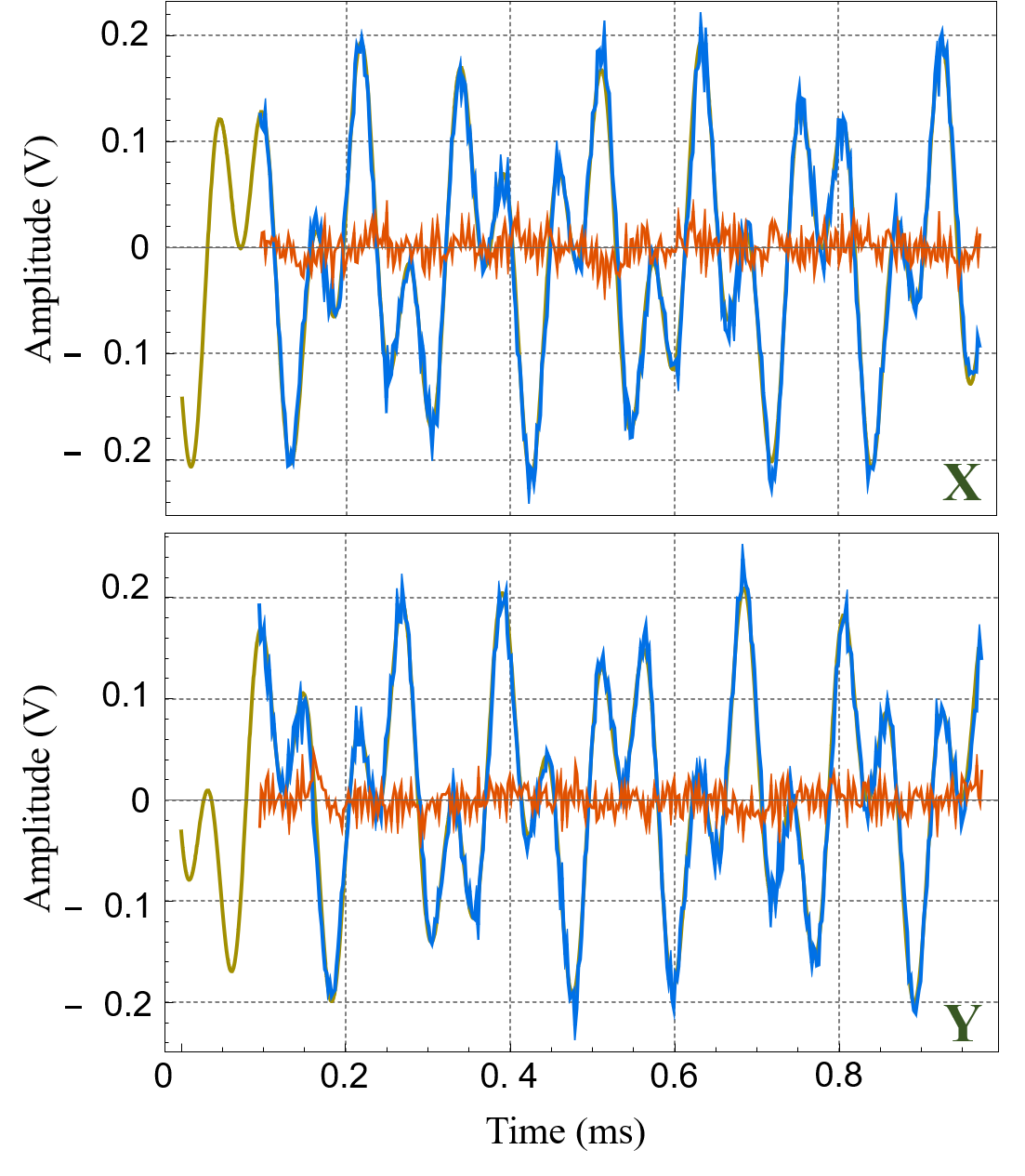}}
\caption{Oscillation signal of the excited membrane mode, measured in the two quadratures at the output of the lock-in amplifier, in the period $(0.1\,\to\,1.0)\,$ms after turning off the pump beam. As a consequence of the two heterodyne detections, i.e., the optical one with a local oscillator 12 kHz apart from the probe field, and the one performed by the lock-in amplifier with a local oscillator at $521.8\,$kHz, the oscillation at $(525.8\,\mathrm{kHz} + f_{\mathrm{m}}) $ splits into two sinusoidal signals at $(8\,\mathrm{kHz} - f_{\mathrm{m}}) $ and $(16\,\mathrm{kHz} + f_{\mathrm{m}})$. The experimental data (blue traces) are obtained from the average over 10 consecutive cycles, and are fitted with the expressions (\ref{eq_XYa}-\ref{eq_XYb}) (yellow lines). Red traces: residuals of the fits.}
\label{fig_XY}
\end{figure}

The lock-in amplifier after the BHD allows to filter out the signal due to the different membrane modes, that are excited in particular when switching on and off the pump beam, due to the steps in the intracavity radiation pressure. As previously mentioned, the lock-in local oscillator is tuned at $(\Wexc - 2\pi \times 4 \,\mathrm{kHz}) $, where for the data shown below $\Wexc = 2\pi \times 525800 \, \mathrm{Hz}$ is chosen to be equal to $\Weff$, measured from the peak of the noise spectrum. As a consequence of the combination of the LO frequency in the BHD and the lock-in local oscillator, the coherent oscillation gives rise to two spectral peaks at the output of the lock-in, centered at 8 kHz and 16 kHz, superposed respectively to the anti-Stokes and Stokes motional sidebands. After turning off the pump beam, the oscillator relaxes. The signal in the two output quadratures of the lock-in can be written as
\begin{eqnarray}
X & = & A \mathrm{e}^{-t/\tau}\,\big\{ \cos\left[2 \pi t (8000 \mathrm{Hz} - f_{\mathrm{m}}) + \phi\right] \nonumber\\
& + & B  \cos\left[2 \pi t (16000 \mathrm{Hz}  + f_{\mathrm{m}}) + \phi+ \delta\phi\right] \big\}
\label{eq_XYa}  \\
Y & = & A \mathrm{e}^{-t/\tau}\,\big\{ \sin\left[2 \pi t (8000 \mathrm{Hz}  - f_{\mathrm{m}}) + \phi\right] \nonumber\\
& - & B  \sin\left[2 \pi t (16000 \mathrm{Hz}  + f_{\mathrm{m}}) + \phi+ \delta\phi\right] \big\} 
\label{eq_XYb}
\end{eqnarray}
where the relative amplitude $B$ and the phase difference $\delta \phi$ between the two sidebands depend on the frequency response of the detection system, and in particular on the filtering function of the lock-in. 
The relaxation time $\tau$ and the oscillation frequency $\Wm$ (in the above expressions, $f_{\mathrm{m}} = (\Wm-\Wexc)/2\pi$) can differ from their natural values due to the optomechanical interaction with the probe field, in case the latter is not perfectly resonant with the optical cavity. An example of the oscillation after the turn off of the pump, measured in the two lock-in quadratures, is shown in Fig. (\ref{fig_XY}) together with the fitting curves obtained by using the expressions (\ref{eq_XYa}) and (\ref{eq_XYb}).

\begin{figure}
\resizebox{0.49\textwidth}{!}{\includegraphics{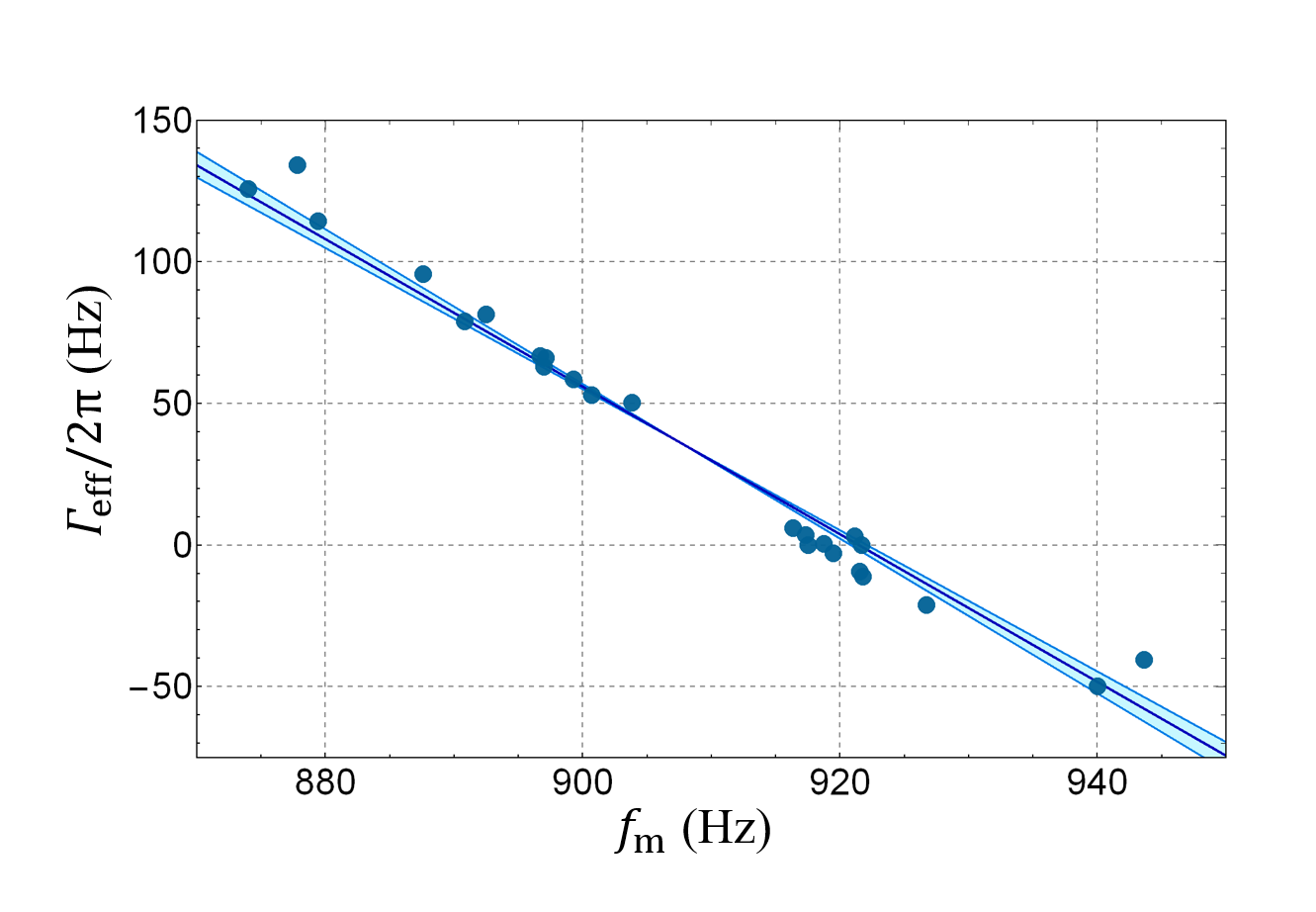}}
\caption{Width of the mechanical resonance, calculated as the inverse of the decay time, as a function of its shift with respect to the excitation frequency. Symbols: experimental data obtained from $50\,$s long time series (each containing 125 experimental cycles lasting $40\,$ms). The probe detuning is varied between time series. The slope of the straight line is calculated from the independently measured cavity width $\kappa$ and mechanical frequency $\Wm$, according to the theoretical relation $\Geff \simeq \delta \Wm\, \times 2 \kappa \Wm/\left[(\kappa/2)^2-\Wm^2\right]$. Its offset is fitted to the experimental data, and the shadowed region accounts for the uncertainty in the measured parameters.}
\label{fig_width_vs_fm}
\end{figure}

The optomechanical effect of the probe field in the present experiment is not negligible. The accuracy in zeroing its detuning is here around $\sim 0.1 \kappa$, and its stability over few minutes is of the same order \cite{Chowdhury2019}. For such a small detuning, the optomechanical frequency shift $\delta \Wm$ and damping $\Gopt=\Geff-\Gm \simeq \Geff$ are roughly proportional, according to $\Gopt \simeq \delta \Wm \frac{ 2 \kappa \Wm}{(\kappa/2)^2-\Wm^2}$ \cite{AspelRMP}. In Fig. (\ref{fig_width_vs_fm}) we report $\Geff/2\pi$, calculated from the measured decay time according to $\Geff = 2/\tau$, as a function of $f_{\mathrm{m}}$. The experimental data points correspond to different time series, acquired by slightly varying the probe locking point. The overall range of detuning is about 20 kHz (corresponding to $0.1 \kappa$). A negative value of the width means that the oscillation amplitude is exponentially increasing (anti-damped oscillator), yet the short (10 ms) measurement time allows to recover the system before a complete cavity unlocking. This analysis allows to select \emph{a posteriori} the time series with the smaller optomechanical effect of the probe. However, in most cases the oscillator damping is dominated by such effect, and the frequency shift of the resonance remains of the order of few Hz (i.e., much larger than the natural width). In order to further reduce the coupling of the probe to the oscillator in our optomechanical system, we envisage to add a further probe, with the wavelength in a region of lower cavity Finesse.  

Following the second proposed protocol, the above described fits on the decaying oscillation have been performed on time series starting $100\,\mu$s after the pump switching off, i.e., when the oscillator is already in the ``classical'' (low purity) regime. At the purpose of searching potential effects of the deformed commutator in the quantum regime, one should quantify a possible shift in the oscillation frequency occuring when the oscillator is still in a high purity state. In order to maintain the highest model independence, it is desirable to avoid making assumptions on the specific dependence of the shift from the purity. Therefore, we limit the analysis to the short time period where the mean phonon number is still approximately constant, as well as the consequent frequency shift. In this work, aiming to demonstrate the application of the proposed protocol, we choose a period of $50\,\mu$s after the pump switching off. Of course, this choice can be refined in the future according to the specific system parameters. Moreover, we limit our analysis to two, $50\,$s long time traces, for which the previous analysis of the decaying oscillation gives a null value of $\Geff$, within the fit uncertainty of less than $1\,$Hz, and therefore a negligible optomechanical effect of the probe laser.

\begin{figure}
\resizebox{0.45\textwidth}{!}{\includegraphics{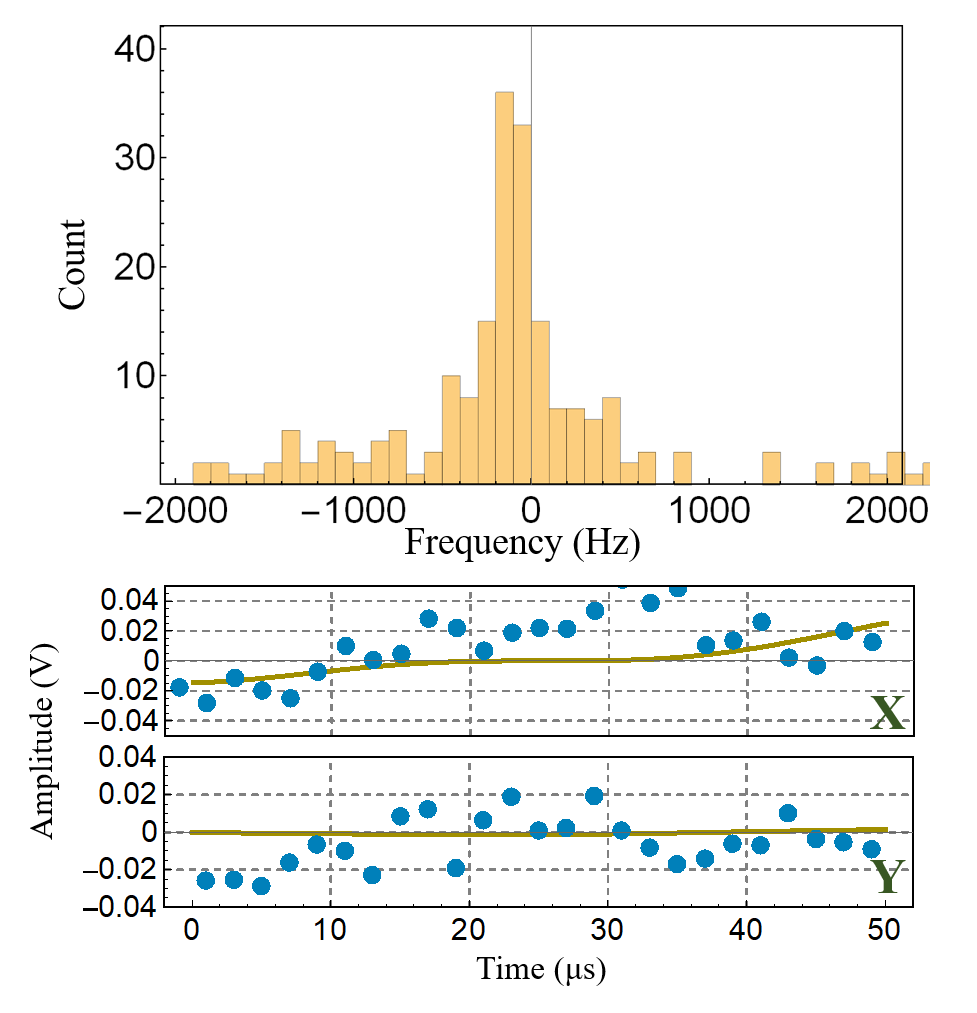}}
\caption{Symbols: residuals obtained by subtracting from the experimental data (averaged over 10 consecutive cycles) the model of Eqs. (\ref{eq_XYa}-\ref{eq_XYb}), with parameters obtained by fitting the data in the $(0.1-1.0)\,$ms interval. Solid lines: fit to such residuals with the model described in the text. Upper panel: histogram of the values of $\dfm^0$ obtained form the analysis of the $X$ quadrature in two, $50\,$s long time series.}
\label{fig_QG}
\end{figure}

We are interested in writing the expression of the quadrature signals including a frequency shift $\dfm$ that rapidly decays, and so small that $\int_0^{\infty}\dfm\,\ud t\,\ll\,1$. Eqs. (\ref{eq_XYa}-\ref{eq_XYb}) should represent its limits, valid after few decay times of $\dfm$. The searched expressions can be obtained by replacing in Eqs. (\ref{eq_XYa}-\ref{eq_XYb})  $\,f_{\mathrm{m}} t\,\to\,f_{\mathrm{m}} t\,+\,\int_0^{t}\dfm\,\ud t'\,-\,\int_0^{\infty}\dfm\,\ud t'$, where the last term accounts for the inclusion of the overall phase shift $2\pi\int_0^{\infty}\dfm\,\ud t$ in the already present phase $\phi$. For short times, the replacement can be written as $\,f_{\mathrm{m}} t\,\to\,\left(f_{\mathrm{m}}+\dfm\right)t \,-\,\int_0^{\infty}\dfm\,\ud t'$, and finally the modified expression $Q_{\dfm}$ for the quadrature $Q\equiv X,Y$ is, at first order in the phase shift, 
\begin{equation}
Q_{\dfm}\,\simeq\,Q\,+\,\frac{\ud Q}{\ud \left(f_{\mathrm{m}} t\right)}\,\times\,\left(\dfm(0)\,t\,-\,\int_0^{\infty}\dfm\,\ud t'\right)
\label{eq_Q}
\end{equation}
where $Q$ is defined in Eqs. (\ref{eq_XYa}-\ref{eq_XYb}). We have fitted with Eq. (\ref{eq_Q}) each of the signals obtained by averaging 10 consecutive, $40\,$ms cycles. More specifically, for each signal, the parameters of $X$ and $Y$ are derived by fitting the data in the $(0.1-1.0)\,$ms interval, as shown in Fig. (\ref{fig_XY}). The extrapolation of Eqs. (\ref{eq_XYa}-\ref{eq_XYb}) to the $(0-50)\,\mu$s region is then subtracted to the experimental data, and the residuals are fitted to the function $ \frac{\ud Q}{\ud \left(f_{\mathrm{m}} t\right)}\,\times\,\left(\dfm^0\,t\,+\,c\right)$ where $\dfm^0$ and $c$ are the two free fitting parameters. In Fig. (\ref{fig_QG}) we show an example of the fitted data, as well as the histogram of the values of $\dfm^0$ obtained from the analysis of the $X$ quadrature (in both the $50\,$s time series considered). This complete set has a mean value of $\langle\,\dfm^0\,\rangle\,=48\,$Hz and a standard deviation of $1390\,$Hz. The analysis of the $Y$ quadrature gives instead a mean value of  $\langle\,\dfm^0\,\rangle\,=30\,$Hz with a standard deviation of $560\,$Hz. Both results are compatible with a null frequency shift.

\section{Conclusions}
\label{conclusion}

The task of conceiving and realizing experiments that could contribute to the search of a physical view joining quantum mechanics and general relativity is, at the same time, extremely useful and difficult, due to the variety and poor concreteness of the different theoretical models approaching this problem. In this framework, it is meaningful to set experimental limits to the parameters characterizing such models even if the comparison between theory and measurable quantities implies additional assumptions that are not straightforward, nor required in the original reasoning. 

An approach that has recently proved effective is based on the effect that a modification of the commutation relation between position and momentum, foreseen by several models, would have on the dynamics of a simple system, such as a harmonic oscillator. Deformed commutators were initially conceived for an ideal point particle. However, in the framework of quantum mechanics it comes natural to extend the application of the commutation rules to any pair of conjugated dynamic variables that describe the motion of the wave function associated with the center of mass of a massive physical system.
Based on this consideration, experimental results obtained with macroscopic oscillators are considered as significant. Nevertheless, the transition from a point particle to a macroscopic physical system may not be obvious. Possible dependence of the deformation parameter of the commutator on macroscopic variables characterizing the oscillator, such as its mass or the number of elementary particles that composes it, has indeed been treated, without questioning the overall validity of the experimental approach \cite{Kumar2020}.

The present work originates from the consideration that the deformed commutator is typically derived as a consequence of a modification of the Heisenberg uncertainty principle, and that in any case it is rooted in the quantum properties of the system.
Therefore, it would not be surprising if peculiar features that are predicted by considering the joined roles of gravity and quantum physics should manifest themselves just on purely quantum objects. Consequently, it is meaningful to carry on the experimental analysis on a system in a high purity state, even exploiting the same measurement principles previously just applied to classical systems.
We have analyzed possible experimental schemes that allow this type of measurement, in particular evidencing a possible effect of the deformed commutator on the oscillation frequency, and described the implementation of one of these schemes. 

Without aiming in this work to validate quantitative results, yet we describe the implementation of the entire protocol. We thus pave the way for an extended data acquisition and analysis campaign that is expected to yield significant results in short terms.

\end{document}